\documentclass[sort&compress,numbers,
article,
aip,
amsmath,amssymb,floatfix,
 reprint,%
]{revtex4-1}

\usepackage{graphicx}
\usepackage{dcolumn}
\usepackage{bm}

\usepackage[utf8]{inputenc}
\usepackage[T1]{fontenc}
\usepackage{mathptmx}
\usepackage{etoolbox}
\usepackage{xcolor}%

\begin{document}


\title{Superionicity in Ammonium Polyhydrides at Extreme Pressures}

\author{K. de Villa}
\email{kyla.devilla@berkeley.edu, Corresponding Author}
\affiliation{Department of Earth and Planetary Science, University of California, Berkeley}
\footnote{Corresponding Author}

\author{X. Wang}
\affiliation{ Department of Chemistry, State University of New York at Buffalo, Buffalo, New York 14260-3000}

\author{E. Zurek}
\affiliation{ Department of Chemistry, State University of New York at Buffalo, Buffalo, New York 14260-3000}

\author{B. Militzer}
\affiliation{Department of Earth and Planetary Science, University of California, Berkeley}
\affiliation{Department of Astronomy, University of California, Berkeley }

\date{\today}

\begin{abstract}
Polyhydrides have been shown to form novel structures at high pressure, which may be found in the interiors of giant planets. With density functional molecular dynamics simulations we studied the behavior of ammonium polyhydride compounds with stoichiometries of NH$_7$, NH$_9$, NH$_{10}$, NH$_{11}$, NH$_{14}$, NH$_{20}$, and NH$_{24}$ which were predicted with crystal structure search methods to be metastable at 100-300~GPa. For every compound, we performed simulations at a range of temperatures (and for several compounds, pressures) covering the solid, superionic and liquid phases. We show that when heated, high pressure ammonium polyhydride compounds exhibit hydrogen superionic diffusion. We demonstrate a number of metrics by which the solid-to-superionic and superionic-to-liquid transitions can be detected from simulation data, including changes in the internal energy and pressure, formation of new chemical species, and atomic diffusion rates. We find that both the solid-to-superionic and the superionic-to-liquid transitions decrease in temperature as proton fraction increases. These trends indicate that above a proton fraction of $\sim$0.97, ammonium hydride structures are likely to directly melt instead of first exhibiting a superionic phase. Our observed melting trend further indicates that at the extreme conditions of ice giant interiors, hydrogen rich ammonium hydrides such as those studied in this work would exist predominantly as liquids rather than exhibiting a superionic phase.
\end{abstract} 

\maketitle


\section{Introduction}

After hydrogen, helium, oxygen, carbon and neon, nitrogen is the sixth most abundant element in our solar system~\cite{asplund2009chemical}, which renders it very likely to find nitrogen bearing compounds in the interiors, surfaces, and atmospheres of different planets. Large quantities of nitrogen are assumed to be stored at high pressure and temperature in the icy mantles of Uranus and Neptune. Presumably, nitrogen was delivered in the form of ammonia, which makes the high pressure properties of hydrogen-nitrogen compounds of interest in planetary science.

Crystal structure prediction (CSP) methods based on \textit{ab initio} electronic structure calculations\cite{hilleke2022tuning} have been very successful in discovering stable polyhydrides with high superconducting transition temperatures ($T_c$), and in the last decade several polyhydride compounds have been synthesized and confirmed experimentally to have superconducting properties\cite{somayazulu2019evidence, drozdov2015conventional,drozdov2019superconductivity,kong2021superconductivity,troyan2021anomalous}. Many polyhydrides observed experimentally or predicted computationally to exhibit high T$_c$ superconductivity use group~I elements as  dopants\cite{zurek2017hydrides}. Song \textit{et al.}\cite{song2019exotic} used CSP methods to explore NH$_n$ (n=1-9) and found that stable phases of NH$_7$ at 25-200~GPa contained NH$_3$ and NH$_4$ units. As NH$_4$ superalkali cations have chemical characteristics similar to alkali metals, we previously investigated the superconducting properties of polyhydrides in which they were substituted for group~I elements in Wang \textit{et al.}\cite{wang2024superconductivity}. We performed CSP searches of NH$_n$ (n=7-16, 20, 24) reporting three novel metastable structures that are metallic and high-symmetry with non-trivial $T_c$s: NH$_{10}$-$Pnma$ (179~K), NH$_{11}$-$Cmc2_1$ (114~K), and NH$_{24}$-$C2$ (134~K) at 300 GPa. In total, 12 novel metastable structures emerged from our CSP search, in addition to the stable NH$_7$-$P4_12_12$ structure previously reported by Song \textit{et al.}.

At high pressures and elevated temperatures, many hydrides are known to exhibit hydrogen superionicity, a mixed solid-liquid phase in which hydrogen ions diffuse like a liquid through a stable lattice of heavier nuclei. In hydrides, the superionic phase has been observed experimentally and computationally for water, ammonia, and ammonia-hydrates\cite{cavazzoni1999,goldman2005bonding,Reinhardt2022,niu2022ultralow,Millot2018,millot2019nanosecond,Gleason2022,cheng2021phase,Prakapenka2021,Kimura2021,Wilson2013,Redmer2011,ninet2012proton,bethkenhagen2013equation,bethkenhagen2015superionic, jiang2017ionic, sugimura2012experimental, kim2022evidence,sun2015phase,weck2022evidence,hernandez2016superionic,ryzhkin1985superionic,matusalem2022plastic,hernandez2023melting,militzer2025ab,forestier2025x,husband2024phase}; for He-H$_2$O, He-NH$_3$, and He-CH$_4$ compounds\cite{kim2022evidence, Liu2019, liu2020plastic, gao2020coexistence}; for various H-C-N, H-C-O, H-N-O, and H-C-N-O compounds\cite{saleh2016novel,conway2021,naumova2021,devilla2023, devilla2025experimental, murayamaUnpublished}; for a broad compositional space of iron, silica, and aluminum alloys relevant to planetary interiors\cite{He2022,gao2022superionic, hou2021superionic,wang2021strong,zhang2022direct, park2023origin,he2025absence, inada2025electrical, park2024electride,zhang2024superionic}; and even for lanthanum polyhydrides\cite{causse2023superionicity}. This unique phase is often harnessed for energy technologies, including for hydrogen storage\cite{prabhukhot2016review}, and solid state batteries\cite{kim2015review} (though these typically employ Li and Na superionic conductors). Superionic phases of H-C-N-O compounds have long been suggested to be influential for dynamo generation in ice giant planets Uranus and Neptune, as these materials exhibit ionic conductivity, often at conditions where they are electronically insulating. During the formation of the ice giant planets, NH$_3$ and H$_2$ were both accreted in large quantities and likely mixed under high pressure conditions. Consequently, investigating the superionic properties of ammonium hydrides under high pressure can provide valuable insights into phase behaviors, transport properties and dynamics of ice giant mantles. 

\begin{figure*}[!ht]
    \centering
    \includegraphics[width=0.8\linewidth]{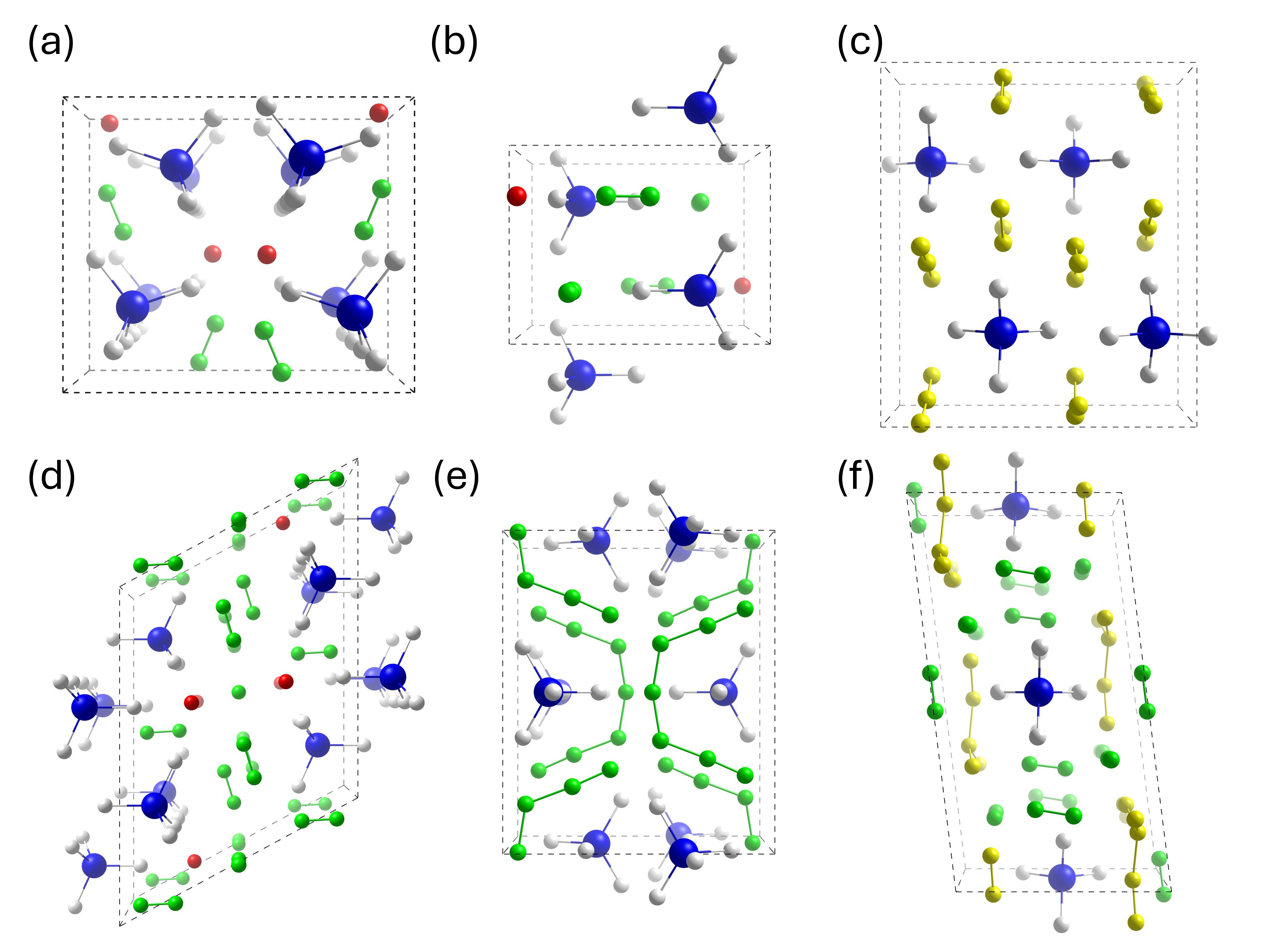}
    \caption{Conventional unit cells of several of the structures considered in this study, with the chemical systems present denoted in brackets. (a) NH$_7$-$Ima2$ [NH$_4^+$, H$_2$, H$^-$], (b) NH$_9$-$P2_1/m$ [NH$_4^+$, 2 H$_2$, H$^-$], (c) NH$_{10}$-$Pnma$ [NH$_4^+$, 2 [H$_3^{0.5-}$]$_\infty$], (d) NH$_{10}$-$C2/m$ [NH$_4^+$, 2.5 H$_2$, H$^-$], (e) NH$_{11}$-$Cmc2_1$ [NH$_4^+$, H$_7^-$], and (f) NH$_{24}$-$C_2$ [NH$_4^+$, 7.5 H$_2$, [H$_5^-$]$_\infty$]. Nitrogen atoms are blue, hydrogen atoms in ammonium are white, H-containing molecular units are green, polymeric hydrogen chains are yellow, and H$^-$ ions are red. The molecular partitioning is highly sensitive to the chosen bond length criterion. In the present work, a cutoff of 1~\r{A} was adopted, consistent with our previous study in which H–H bonds with $-$ICOHP values exceeding 2~eV per bond were regarded as significant~\cite{wang2024superconductivity}. The remaining structures studied (NH$_7$-$P4_12_12$, NH$_9$-$Cmcm$, NH$_9$-$P2_1$, NH$_9$-$Pmnm$, NH$_{14}$-$P$-1-I, NH$_{14}$-$P$-1-II, and NH$_{20}$-$P$-1) are shown in Supplemental Fig.~1.}
    \label{fig:str-xy}
\end{figure*}

Building on our previous predictions of metastable NH$_n$ phases with exceptionally high hydrogen concentrations\cite{wang2024superconductivity}, this work explores the possibility of superionic behavior in these hydrogen rich compounds. We perform density functional molecular dynamics (DFT-MD) simulations for each predicted structure at a range of temperatures, and for selected compounds, a range of pressures, to assess the possibility of a superionic phase. Materials had different initial pressures at 0~K: 100~GPa structures include NH$_7$-$P4_12_12$, NH$_9$-$Pmnm$, NH$_9$-$Cmcm$, and NH$_{10}$-$C2/m$; and 300~GPa structures include NH$_9$-$P2_1$, NH$_{10}$-$Pnma$, NH$_{11}$-$Cmc2_1$, NH$_{14}$-$P$-1-I, NH$_{14}$-$P$-1-I, NH$_{20}$-$P$-1, and NH$_{24}$-$C2$. NH$_7$-$Ima2$ was simulated at 100 and 300~GPa, and NH$_9$-$P2_1/m$ was simulated at 100, 200, 300, and 400~GPa. Depictions of the unit cells of these structures are given in Fig.~\ref{fig:str-xy} and Fig.~S1. As Ref.\cite{wang2024superconductivity} focused on superconducting phases, only those that were metallic were previously discussed. However, in this manuscript we are concerned with the proton dynamics, therefore all metastable phases previously predicted are considered.

We examine the common and divergent features of each structure, finding that proton fraction (n$_{\rm H}$/n$_{\rm total}$) strongly influences temperature dependent bonding motifs, species lifetimes, and pressure and internal energy trends. Remarkably, all structures studied exhibit superionic phases upon sufficient heating, demonstrating that the superionic state remains stable even at high proton fractions in ammonium polyhydrides. Strong correlations emerge between proton fraction and diffusion behaviors, with both solid-to-superionic and superionic-to-liquid transitions decreasing in temperature as proton fraction increases. However, all structures melt below the isentropes of Uranus and Neptune, indicating that such high hydrogen concentrations in ammonium polyhydrides likely prevent superionic behavior in ice giant interiors.

\section{Methods}

\subsection{DFT-MD Simulations}
Electronic structure calculations were performed with the Perdew, Burke, and Ernzerhof (PBE) exchange-correlation functional~\cite{perdew_generalized_1996} as implemented in the Vienna \textit{Ab Initio} Simulation Package (VASP)~\cite{kresse_efficient_1996}. The projector-augmented wave (PAW) method~\cite{blochl_projector_1994} was used to treat the core states, with core radii of 0.8 and 1.1~\AA~for hydrogen and nitrogen, respectively. Valence orbitals were represented by plane waves up to a cutoff energy 1100~eV. The Brillouin zone was sampled with either the $\Gamma$-point only or with a 2$\times$2$\times$2 $k$-point grid, which was determined based on convergence criteria of 1\% in energy, pressure, and uniaxial components of the stress tensor at 0~K and at 500~K using a 3$\times$3$\times$3 $k$-point grid. Electronic states were populated according to a Fermi-Dirac distribution using the Mermin functional~\cite{mermin_thermal_1965}, which incorporated excited electronic states at finite temperature within the Kohn-Sham formalism~\cite{kohn_self-consistent_1965}.

Crystal structures were obtained from Refs.\cite{wang2024superconductivity,song2019exotic}. Focus is placed on the 13 metastable systems in Fig.~\ref{fig:str-xy} and Fig. S1 because their formation enthalpies lie within 60 meV atom$^{-1}$ of the convex hull, which is set by the enthalpy of the stable structures NH$_3$\cite{pickard_2008_nh3}, NH$_7$-$P4_12_12$\cite{song2019exotic}, and H$_2$\cite{pickard_2007_h2}, and were chosen because this falls within the 90th percentile of DFT-calculated metastability\cite{sun2016thermodynamic}. Although at finite temperature it is free energy, not enthalpy, that controls structural stability, we expect that the low temperatures at which we study structural properties (such as the onset of superioncity) will not significantly influence the stability of these structures. Structural parameters for NH$_7$-$P4_12_12$ which were previously published by \citet{song2019exotic}; and for NH$_{10}$-$Pnma$, NH$_{11}$-$Cmc2_1$, and NH$_{24}$-$C2$ by \citet{wang2024superconductivity} are provided in the supplementary material of their respective publications. Structural parameters for all other crystals are given in Supplemental Information.

Superionic diffusion was studied with density functional theory molecular dynamics simulations. Supercells containing 80 to 168 atoms were simulated in the $NVT$ ensemble equilibrated with the Nos\'e-Hoover thermostat~\cite{nose_unified_1984,nose_constant_1991}, using a 0.5~fs timestep.  Supercells were heated at fixed density in increments of 50~K from 100~K to 1000~K, and in increments of 250~K from 1000~K to their melting point, as determined using the heat-until-it-melts approach~\cite{HUM}. At each temperature, the system was allowed to equilibrate for at least 1~ps before the system was heated to the subequent temperature, with velocities scaled between temperatures to accelerate equilibration. Constant temperature simulations were performed from 1.5 to 100~ps, with timescales varying between simulations to ensure that phases at each temperature were accurately discerned.

Melting temperatures calculated here are merely an upper bound as this method can result in superheating, particularly for simulations with relatively few atoms. More sophisticated methods, such as coexistence simulations\cite{smallenburg2024simple}, the Z method\cite{wang2013modified}, and thermodynamic integration\cite{gonzalez2023ab} may alternatively be applied to obtain more accurate melting lines, typically with an error of 500~K or less for H-C-N-O materials\cite{devilla2023,murayama2025review}, though no comprehensive comparison as a function of pressure and composition has been completed for this chemical space. Superionic transitions are known to have minimal thermal hysteresis\cite{freiheit2001order} and we estimate the errors on superionic diffusion temperatures are at most 50~K, the size of the temperature increments used.

\subsection{Cluster Analysis}
For each chemical structure, temperature, and pressure simulated, we analyzed the prevalent chemical. Following the heuristics 
employed by Goldman \textit{et al.}\cite{goldman2005bonding,goldman2006first,goldman2010synthesis}, we define clusters based on bond length cutoffs, and categorize these species as stable or as transition states based on their lifetime. While maximum bond lengths are somewhat ambiguous in condensed phases, especially in materials with hydrogen bonding, here we infer the maximum bond length from the first minimum in the radial distribution function. We set cutoff radii to 0.9~\AA~for H-H bonds and 1.4~\AA~for H-N bonds. The resulting cluster classification was particularly sensitive to the H-H bond cutoff because large polymeric networks were identified at higher pressures if the H-H cutoff was set to 1.0~\AA. We observe that N-N distances are always greater than 1.8~\AA, with a first minimum occurring around 3~\AA, indicating that no traditional bonding takes place between N atoms in these hydrogen rich systems.

We make the approximation that for a species to be stable and identified experimentally, its vibrational modes must be observable\cite{aroca1982molecular}. As a result, we require that bonds persist for at least two vibrational stretching periods. In crystalline ammonia at high pressures\cite{ninet2014experimental}, H-N vibrational modes are centered around 3500~cm$^{-1}$, corresponding to a period of 9.5~fs, thus we set the minimum bond lifetime for H-N to 20~fs. High pressure H-H bonds center around 4400~cm$^{-1}$ or 7.6~fs\cite{zong2020understanding}, so we similarly set the minimum bond lifetime for H-H to 20~fs. N-N vibrational frequencies center around 2350~cm$^{-1}$ or 14.2~fs from 1-6~GPa\cite{mccluskey2002n}, but have been shown to increase slowly with pressure\cite{moore1995vibrational}. Therefore, we use a minimum lifetime of 50~fs which is roughly equal to 3.5 vibrational stretching periods for N-N bonds.


\section{Results and Discussion}

\subsection{Determining Solid-to-Superionic Transition Temperature}

\begin{table*}
  \centering
  \caption{All structures studied in this work. For each structure we list the stoichiometry, space group, pressure at 0~K, constituent units in the cell at 0~K, metallicity, solid-to-superionic transition temperature (T$_{SI}$) and superionic-to-liquid (melting) transition temperature (T$_m$) as inferred from DFT-MD simulations.}
  \label{tab:example}
 \begin{tabular}{c|c|c|c|c|c|c}

    Stoichiometry & Space Group & Pressure at 0~K (GPa) & Constituent Units & Metallic? & T$_{SI}$ (K) & T$_m$ (K) \\
    \hline
    NH$_7$ & $Ima2$ & 100 & NH$_4^+$, H$_2$, H$^-$ & No & 700 & 2500 \\

     & & 300 &  && 600 & 3000 \\
\hline
    NH$_7$ & $P4_12_12$ & 100 & NH$_4^+$, H$_2$, H$^-$ & No & 700 & 2750 \\
    \hline
    NH$_9$ & $Cmcm$ & 100 & NH$_4^+$, 2 H$_2$, H$^-$ & No &  600 & 2000\\
\hline
    NH$_9$ & $P2_1$ & 300 & NH$_3$, 3 H$_2$ & No &  400 & 2500 \\
\hline
    NH$_9$ & $P2_1/m$ & 100 & NH$_4^+$, 2 H$_2$, H$^-$& No &  550 & 1500 \\

     & & 200 &&& 550 & 2000 \\

     &  & 300 &&& 500 & 2250 \\

    && 400 &&& 400 & 2250 \\
 \hline
    NH$_9$ & $Pmnm$ & 100 & NH$_4^+$, 2 H$_2$, H$^-$ & No &  600 & 1750 \\
    \hline
    NH$_{10}$ & $C2/m$ & 100 & NH$_4^+$, 2.5 H$_2$, H$^-$ & No &  600 & 1750 \\
    \hline

    NH$_{10}$ & $Pnma$ & 300 & NH$_4^+$, 2 [H$_3^{0.5-}$]$_\infty$ & Yes &  450 & 2250 \\
    \hline
    NH$_{11}$ & $Cmc2_1$ & 300 & NH$_4^+$, H$_7^-$ & Yes &  500 & 1750 \\
    \hline
    NH$_{14}$ & $P$-1-I & 300 & NH$_4^+$, 3 H$_2$, H$_4^-$ & No &  600 & 1250 \\
     \hline

    NH$_{14}$ & $P$-1-II & 300 & NH$_4^+$, 2.5 H$_2$, H$_4$, H$^-$ & No &  500 & 850 \\
    \hline
    NH$_{20}$ & $P$-1 & 300 & NH$_4^+$, 2.5 H$_2$, 0.5 H$_8$, H$_7^-$ & No &  400 & 750 \\
    \hline
    NH$_{24}$ & $C2$ & 300 & NH$_4^+$, 7.5 H$_2$, [H$_5^-$]$_\infty$ & Yes &  350 & 850 \\
\end{tabular}
\end{table*}

The superionic phase is characterized by the liquid-like diffusion of small ions through a stable crystalline lattice of bigger ions. For the case of ammonium hydrides, superionic behavior occurs as H ions diffuse through stable N ion sublattices. Superionic diffusion is known to occur for NH compounds: NH$_3$ superionicity has been observed experimentally and computationally above $\sim$50~GPa\cite{hernandez2016superionic}; and superionic phases of NH$_7$ have been observed computationally at pressures above $\sim$25~GPa\cite{song2019exotic}. The $R$-$3m$ and $P4_12_12$ phases of NH$_7$ were predicted to be the lowest point on the NH$_3$-H$_2$ convex hull from 25-60 and 60-200~GPa, respectively.

\begin{figure*}[h]
    \centering
    \includegraphics[width=\textwidth]{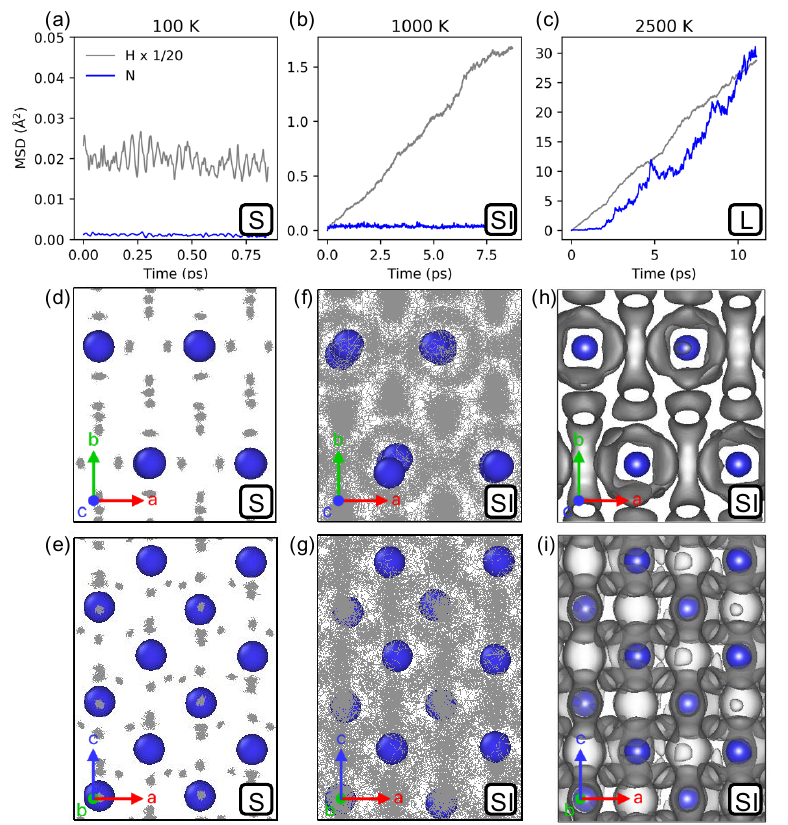}
    \caption{Visualization of superionic diffusion in the NH$_{10}$-$Pnma$ structure. The phase is labeled in the lower right corner of each panel: S for solid, SI for superionic, and L for liquid. (a-c) Mean squared displacement (MSD) plots for H and N ions in the solid phase (100~K), the superionic phase (1000~K), and the liquid phase (2500~K). The H MSD has been divided by 20 in all panels for ease of comparison with N. (d-e) Trajectory plots for the solid phase looking at two different faces. Blue spheres show N, and gray lines trace the motion of H ions. For the solid phase, we see H ions are well constrained in their potential wells. (f-g) Trajectory plots for the superionic phase. Gray lines now show the pathways of H ions as they rotate in NH$_4^+$ and diffuse through the cell. (h-i) Isosurfaces of the time averaged H positions match well with the trajectory plots in (f-g), showing that H ions spend much of their time bound in rotating NH$_4^+$ ions, and diffusing through 
    interconnected pathways surrounding the N ions. }
    \label{fig:msd_etc}
\end{figure*}

The traditional indication that a material has transitioned from the solid phase to the superionic phase is the onset of H ion diffusion, which can be characterized statistically through calculation of the mean squared displacement (MSD) of H ions, defined as  
\begin{equation}
{\rm MSD}(t)\equiv\left<(\mathbf r_i(t)-\mathbf r_i(0))^2\right>
\end{equation}
where $\mathbf r_i(t)$ is the position of ion $i$ at a time $t$, $\mathbf r_i(0)$ is its initial position, and the MSD at each time is averaged over all ions of a given atomic species. 

We first discuss the results for NH$_{10}$-$Pnma$, as this structure was previously studied extensively as a potential superconductor\cite{wang2024superconductivity}. As shown in Fig.~\ref{fig:msd_etc}(a-c), the solid phase exhibits a flat yet oscillating MSD for all ion types, indicating that ions are vibrating in their potential wells, but they cannot diffuse. The superionic phase is characterized by a linearly increasing MSD for H ions and a flat MSD for all other ion types, the hallmark of this mixed solid-liquid phase. The liquid phase exhibits linearly increasing MSDs for all ion types, with lighter ions exhibiting a greater MSD slope due to their decreased mass resulting in them moving faster in thermodynamic equilibrium, with $\frac{1}{2} mv^2 \sim k_BT$, corresponding to a more rapid diffusion speed. For the system shown in Fig.~\ref{fig:msd_etc}, the slope of the H MSD is approximately 20 times that of the N MSD, which is similar to the ratio of their masses, 14. This is consistent with the diffusion rates in gases, which scale as MSD~$\sim r^2 \sim v^2 \sim m^{-1}$.

Superionic diffusion may also be characterized visually by the hopping of protons out of their initial positions into local, energetically accessible locations, as shown in Fig.~\ref{fig:msd_etc}(d-g). For the solid phase we again observe H ions, whose positions over time are tracked by gray lines, oscillating in their potential wells, but not diffusing. Upon the transition to the superionic phase, H ions diffuse widely throughout the cell. Another method for visual determination of superionic diffusion is with H ion density isosurfaces, as shown in Fig.~\ref{fig:msd_etc}(h-i). By tracking the frequency of H ion occupation in each cubit of the simulation cell, we characterize regions most likely for H ions to occupy. For the NH$_{10}$-$Pnma$ structure, many H ions are bound into rotating NH$_4^+$ ions, with many others diffusing in pathways parallel to the c-axis. 

While the aforementioned determinations of superionicity are sufficient in most cases, these metrics often fail when trying to determine the solid-to-superionic transition in materials that diffuse at comparatively low temperatures, as we observe in this study. One complication is at temperatures between the solid and the superionic phases, many compounds enter an ionically excited or plastic state\cite{ninet2012proton,liu2020plastic}, in which species such as NH$_4^+$ ions may begin to rotate, or H ions may be able to hop between two adjacent sites, but not through the entire cell. This is similar to how carbonate ions, CO$_3^{-2}$, can rotate in CaCO$_3$ crystal structures without triggering any diffusion of oxygen ions\cite{li2017determination}. Excited states exhibit a brief increase in MSD followed by a flattening to a non-zero value. However, we find that at low temperatures at which we observe ammonium hydride superionicity, this MSD flattening effect can take upwards of 50-100~ps, a timescale that is inefficient and often inaccessible with traditional DFT-MD methods.

\begin{figure}
    \centering    \includegraphics[width=1\linewidth]{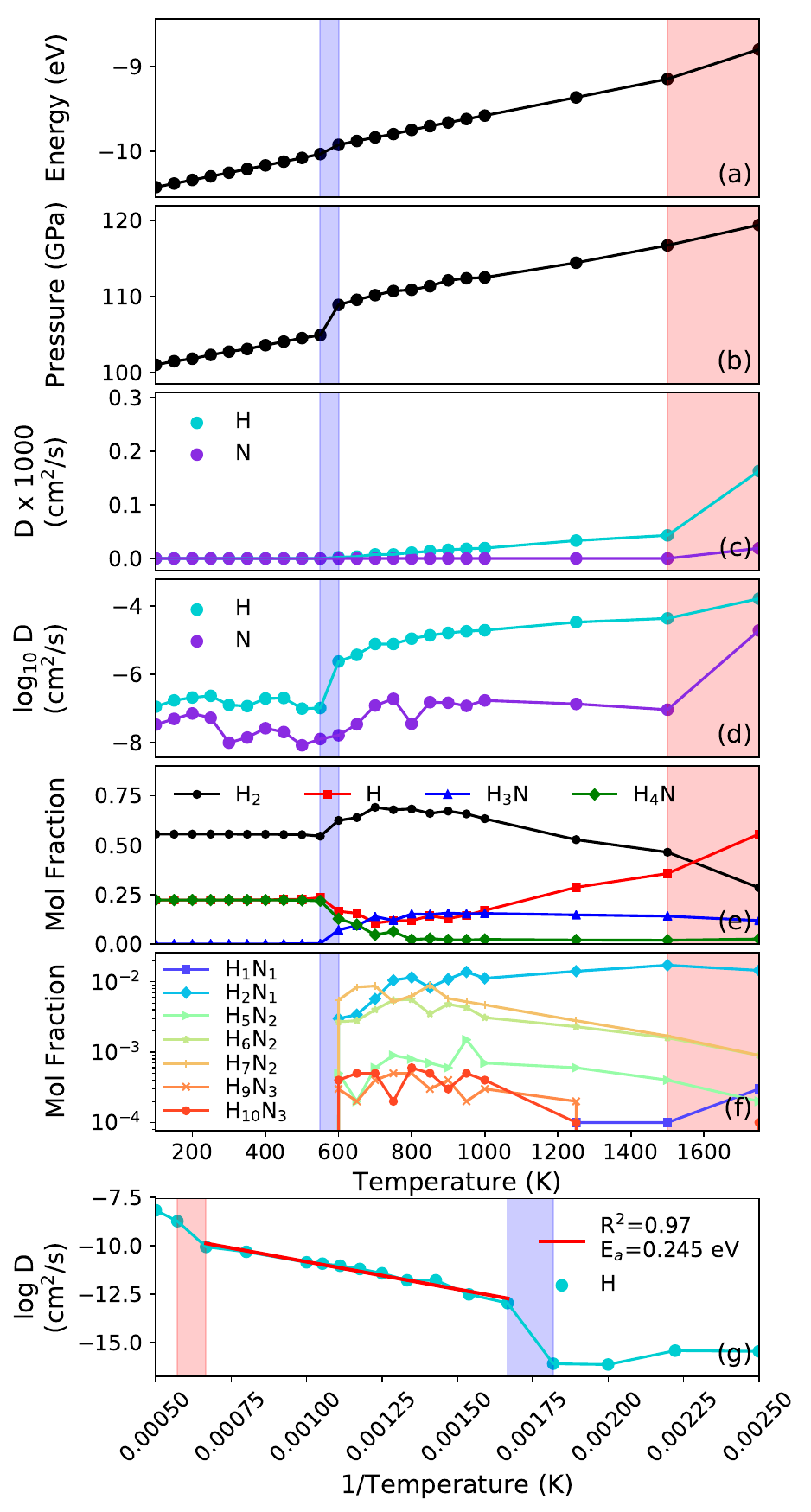}
    \caption{A series of metrics by which the solid-to-superionic transition temperature may be determined using simulation results of the NH$_{10}$-$C2/m$ structure at 1.43~g/cm$^3$ as an example. Blue and red shaded bars indicate the temperature range of the solid-to-superionic transition and the superionic-to-liquid transition, respectively. (a-b) When performing DFT-MD simulations in 50~K increments, small jumps in energy and pressure are often observed at the superionic transition. (c-d) A jump in H ion diffusivity (abbreviated D) is the typical signature of superionicity, though this is best visualized on log or log$_{10}$ scale. (e-f) We find that upon the transition to the superionic state, the concentrations of existing H-bearing species change as H ions diffuse and form new bonds. Similarly, new species are formed, which facilitate H ion diffusion through proton exchange. (g) Superionic hydrogen diffusion typically follows an Arrhenius relationship, where the natural log of diffusivity (calculated from the MSD) follows linearly with 1/T. A high R$^2$ value in this least-squares fit indicates the temperature regime of superionicity has been successfully identified.}
    \label{fig:SI_methods}
\end{figure}

In Fig.~\ref{fig:SI_methods}, we demonstrate a number of alternative metrics to characterize the solid-to-superionic transition using simulation results, now using as an example the NH$_{10}$-$C2/m$ structure at $\sim$100~GPa. Blue and red shaded bars indicate the temperature range of the solid-to-superionic and superionic-to-liquid transitions. Small shifts in (a) internal energy (details of our finite temperature DFT calculations are given in section 3a) and (b) pressure often occur as a material becomes superionic due to H ions escaping the potential wells of their ground state positions, potentially accompanied by a slight rearrangement or volume change of the heavy ion lattice. 

Although distinguishing excited states from superionic states can be difficult by looking at MSDs alone, comparing temperature-dependent self-diffusion coefficients for H ions, also called diffusivities, can often reveal the onset of superionicity. Here, diffusivity is calculated from the MSD:
\begin{equation}
{\rm D}=\lim_{t \to \infty} \frac{\text{MSD(t)}}{6t}
\end{equation}
where $t$ is the MD simulation time, although diffusivity may be equivalently calculated using velocity auto-correlation functions\cite{AT87}. Fig.~\ref{fig:SI_methods}(c-d) show that H ion diffusion coefficients, particularly when plotted on a log scale, will show a sharp discontinuity between the solid or excited phase and the superionic phase. While the traditional diffusivity benchmark for the onset to the superionic phase in ices has been taken as $\sim$10$^{-4}$ following~\citet{cavazzoni1999}, we find that a more appropriate threshold is $\sim$10$^{-6}$. The combination of linear and log scale diffusivity plots also reveals the superionic-to-liquid transition, as both H and N ions exhibit a jump in diffusivity. 

Superionic diffusion is typically well represented by an Arrhenius relationship, which assumes diffusion is controlled by thermally activated jumps over energy barriers that have very similar activation energies, $E_a$. In this case, the natural log of the hydrogen ion diffusivity is a linear function of the inverse of temperature. The activation energy can be calculated from the negative of the slope of this line\cite{gao2020classical}:
\begin{equation}
    \ln(D) = \ln(D_0) - \frac{E_a}{k_B T} .
    \label{eq:Arr}
\end{equation}
Using hydrogen diffusivities calculated at discrete temperatures, one can perform a least-squares fit with $1/T$. A high R$^2$ value of this fit indicates that the temperature regime of superionicity has been identified correctly and the Arrhenius relationship applies. Similarly to (d), jumps in H ion diffusivity are also quite apparent at both the solid-to-superionic and superionic-to-liquid transitions. 

Finally, analysis of chemical species present in the simulation cell at different temperatures is also a promising indicator of the solid-to-superionic transition. Fig.~\ref{fig:SI_methods}(e-f) show that before the solid-to-superionic transition, the concentrations of major constituents H, H$_2$, NH$_3$, and NH$_4$ are relatively stable, and no other species are present. However, upon the onset of superionic diffusion, a number of other species can form, even if only briefly. At temperatures closest to the solid-to-superionic transition, small concentrations of NH$_3$ and NH$_2$ appear, coincident with a decrease in concentration of NH$_4$, as H ions can diffuse away from NH$_4$ clusters. At low temperatures, larger transition states with 2-3 N ions also form, facilitating proton exchange. As the temperature continues to increase, these large complexes are no longer energetically favorable, and smaller species like HN appear. Note that cluster analysis only involves geometric relationships between atoms, and does not take advantage of any other information such as electronic structure, resulting in us not assigning charges to the aforementioned chemical species, though many must exist in ionic rather than molecular states. 


\subsection{Phase Diagram of NH$_9$}

\begin{figure}
    \centering
    \includegraphics[width=1\linewidth]{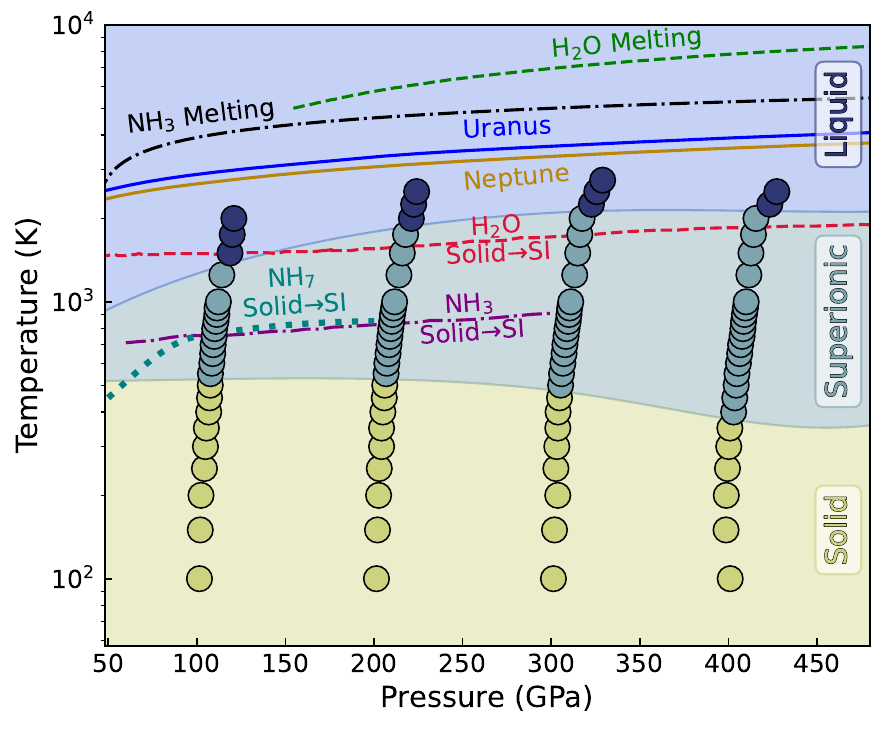}
    \caption{The phase diagram of the NH$_9$-$P2_1/m$ structure from 100--400~GPa showing the solid, superionic, and liquid phases. Also plotted are the melting and solid-to-superionic transitions of water\cite{Wilson2013} and ammonia\cite{hernandez2023melting}, and the solid-to-superionic transition of NH$_7$ ($R-3m$ from 20-60~GPa and $P4_12_12$ from 60-200~GPa\cite{song2019exotic}). These phase transitions are compared to the isentropes of Uranus and Neptune\cite{militzer2025ab}. Because of its higher proton fraction, the NH$_9$-$P2_1/m$ structure becomes superionic and melts at lower temperatures than NH$_3$ and NH$_7$ compounds.}
    \label{fig:nh9phasediagram}
\end{figure}

We performed DFT-MD simulations of the NH$_9$-$P2_1/m$ structure on a density-temperature grid to assess the diffusive behavior from 100-400~GPa and 100-2500~K. Fig.~\ref{fig:nh9phasediagram} shows the resulting phase diagram. This compound exhibits a lower temperature solid-to-superionic transition than NH$_3$, NH$_7$, and H$_2$O. Although less proton rich compounds such as H$_2$O and NH$_3$ likely exist in their superionic phases in Uranus and Neptune, we can infer from this phase diagram that proton rich ammonium hydrides such as the NH$_9$-$P2_1/m$ structure instead exist as liquids at the pressures and temperatures of ice giant interiors, even if one adopts the colder temperature profiles that were recently derived in Ref.~\cite{militzer2025ab}. 

\subsection{Pressure and Energy Changes During Phase Transitions}

\begin{figure}
    \centering
    \includegraphics[width=1.1\linewidth]{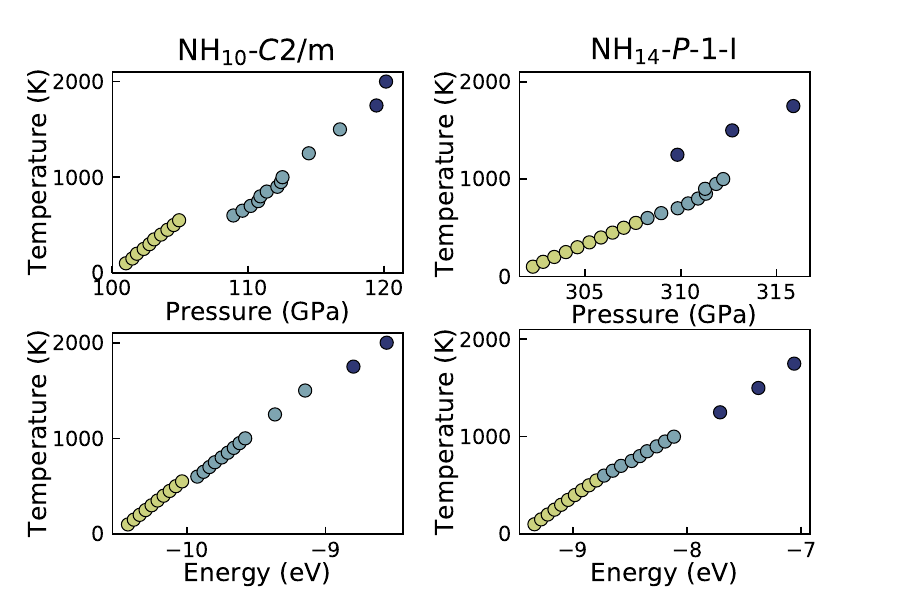}
    \caption{Pressure-temperature and energy-temperature trends for the NH$_{10}$-$C2/m$ and NH$_{14}$-$P$-1-I structures. Yellow, light blue, and dark blue markers correspond to the solid, superionic, and liquid phases, respectively. For the NH$_{10}$-$C2/m$ structure, we see increases in pressure and energy at the solid-to-superionic and superionic-to-liquid transitions. 
    For the NH$_{14}$-$P$-1-I structure, we observe no such increases but the pressure decreases upon melting, which means the liquid is denser than the superionic phase, and the melting line has a negative Clapeyron slope, $dT_m/dP<0$.}
    \label{fig:twophasediagram}
\end{figure}

The NH$_7$-NH$_{11}$ structures simulated typically followed similar pressure and temperature trends when heated at constant volume to those shown in Fig.~\ref{fig:SI_methods}(a-b) and Fig.~\ref{fig:nh9phasediagram}. In these less proton rich structures, small increases in pressure and energy observed were upon the solid-to-superionic and superionic-to-liquid transitions. However, more proton rich structures NH$_{14}$-NH$_{24}$ show no such energy discontinuities upon phase changes, and instead exhibit a pressure decrease upon melting, which means the liquid is denser than the superionic phase, and the melting line has a negative Clapeyron slope, $dT_m/dP<0$. Fig.~\ref{fig:twophasediagram} demonstrates these differences in transition behaviors using structures NH$_{10}$-$C2/m$ and NH$_{14}$-$P$-1-I as examples. This phenomenon is known to occur in water upon melting at ambient pressure, resulting in solid water (ice) floating on top of the liquid phase. A negative Clapeyron slope has also been observed upon melting in doubly superionic CH$_2$N$_2$\cite{devilla2023} at $\sim$500~GPa, and in plastic-to-superionic transitions for He-NH$_3$ compounds\cite{liu2020plastic} at $\sim$5-50~GPa. 


\subsection{Superionic Diffusion and Melting Trends with Proton Fraction}
\begin{figure}[h!]
    \centering
    \includegraphics[width=1\linewidth]{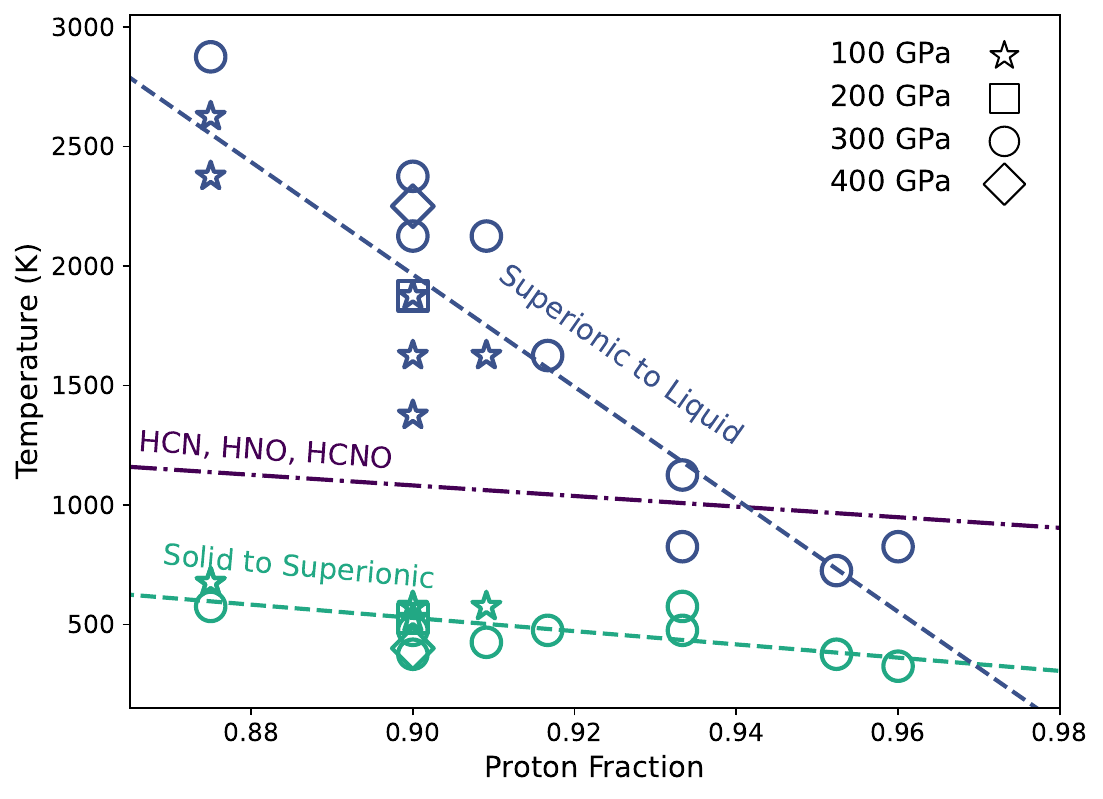}
    \caption{The solid-to-superionic and superionic-to-liquid phase transition temperatures are plotted for different pressures and compounds as function of hydrogen fraction. The linear fits to the two transition temperatures show a decreasing trend for structures that contains more hydrogen atoms because they form weaker chemical bonds. The solid-to-superionic transition temperature of HCN, HNO and HCNO compounds reported by de Villa \textit{et al.}\cite{devilla2023} is approximately twice as high as we report here for NH compounds, emphasizing that the atoms in these NH materials are not bonded as strongly.}
    \label{fig:SI_MELT_TRENDS}
\end{figure}

Over a pressure interval from 100 to 400~GPa, we have simulated 13 different ammonium polyhydrides with proton fractions ranging from 0.875 to 0.96. Using the methods mentioned above, we determined the temperatures of solid-to-superionic and superionic-to-liquid transition for different pressures, as listed in Table 1. In Fig.~\ref{fig:SI_MELT_TRENDS}, we plot the transition temperatures of different materials against their proton fractions. The linear fits to the two transition temperatures show a decreasing trend with increasing proton fraction, indicating that a greater proton fraction correlates to weaker chemical bonds. Also shown is the linear fit of solid-to-superionic transition temperatures for HCN, HNO, and HCNO compounds provided in de~Villa \textit{et al.} 2023\cite{devilla2023}, which shows transition temperatures approximately twice as high as we report here for NH compounds, emphasizing the weaker chemical bonds present in these hydrogen rich materials. We observe that higher pressure structures of the same proton fraction typically melt at higher temperatures - for example, for NH$_9$ structures (proton fraction=0.9), 100-200~GPa structures melt between 1250 and 2000~K, whereas 300-400~GPa structures melt between 2000 and 2500~K. This is consistent with the behavior of well-studied planetary ices such as water and ammonia\cite{Wilson2013,militzer2025ab,hernandez2023melting}. Conversely, higher pressure structures of the same proton fraction appear to exhibit lower temperature superionic diffusion, though only by 50-100~K. Notably, the slope of the superionic-to-liquid transition line is much steeper than the solid-to-superionic transition line. This means that as proton fraction increases, materials exhibits superionic diffusion over a narrower temperature range. At a proton fraction of $\sim$0.97, the linear fit lines of the solid-to-superionic and the superionic-to-liquid transitions cross. This likely indicates that for high proton fractions, a structure melts directly because the interactions between N ions are too weak for a superionic phase to be stable. 

\begin{figure*}[h!]
    \centering
    \includegraphics[width=0.9\linewidth]{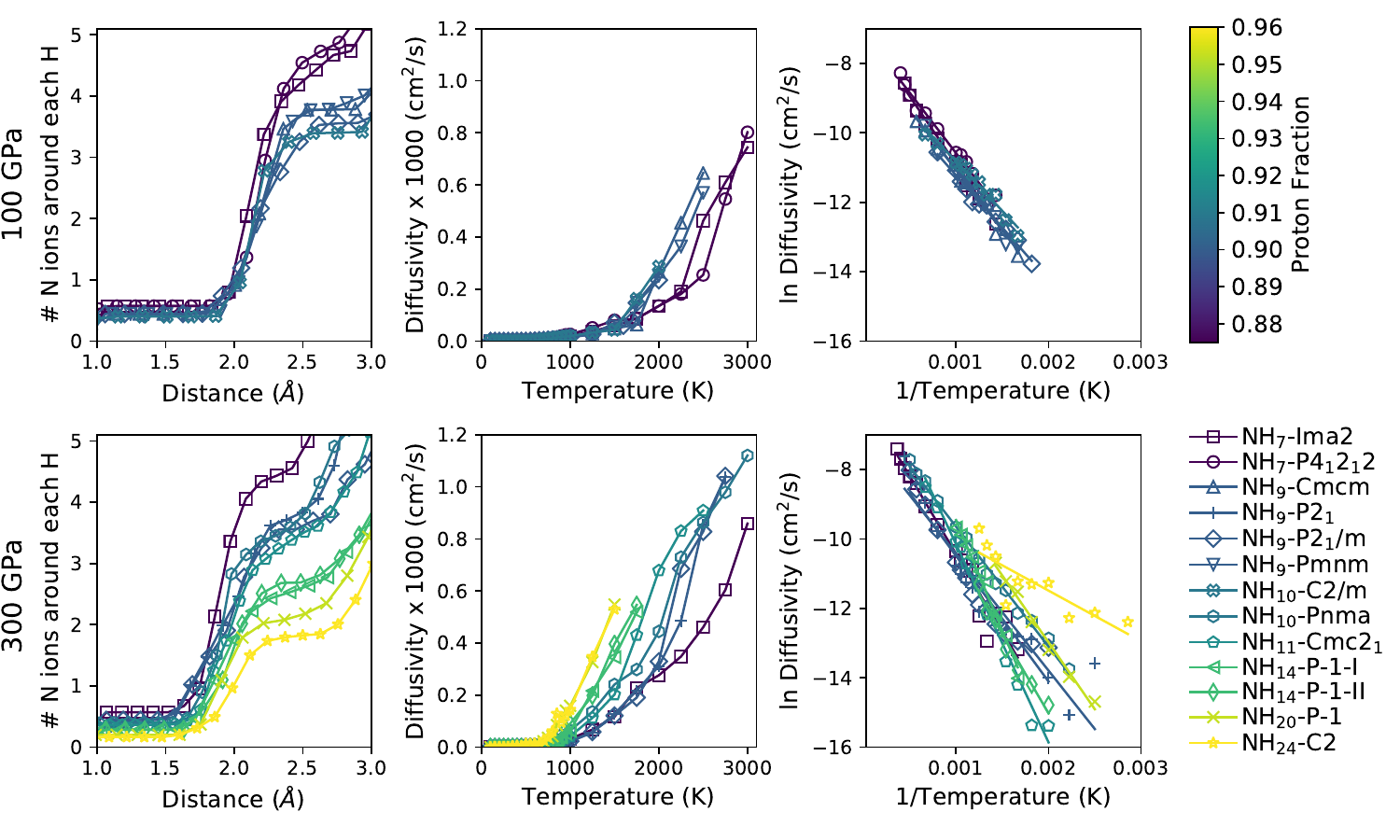}
    \caption{Structural and diffusivity trends for NH$_7$-NH$_{24}$ structures at 100~GPa (top) and 300~GPa (bottom). (left) Coordination plot for H: average number of N ions within a given distance of an H ion. (middle) H ion diffusivities on a linear scale for all temperatures simulated. (right) H ion diffusivities for the superionic phase plotted in Arrhenius trends. The slopes of the fitted, straight lines approximately represent the heights of energy barriers in the H ion diffusion pathways.}
    \label{fig:nOfR_etc}
\end{figure*}

The influence of proton fraction on solid-to-superionic and superionic-to-liquid transition temperatures is dependent on the bonding environments in each structure. Fig.~\ref{fig:nOfR_etc} relates H coordination number to diffusive behaviors for structures with stoichiometries of NH$_7$-NH$_{24}$, with the top row plotting simulation results from 100~GPa, and the bottom row for 300~GPa, and with trend lines colored according to the proton fraction of each structure. The left panels show the average number of N ions within a certain distance of a given H ion. While there is a modest increase in the number of nearby N ions as pressure increases from 100 to 300~GPa, there is a pronounced decrease in the number of nearby N ions as proton fraction increases. At 300~GPa, the NH$_7$-$Ima2$ structure (proton fraction=0.875) has $\sim$5 N ions within 2.5~\AA, whereas the NH$_{24}$-$C2$ structure (proton fraction 0.96) has less than 2 N ions within the same distance. The greater concentration of H ions in more proton rich structures like NH$_{24}$-$C2$ results in each H ion having fewer N ions to bond with.

The middle panels of Fig.~\ref{fig:nOfR_etc} show the H ion diffusivities for 100 and 300~GPa. Again we see the influence of proton fraction on diffusion, with the more proton rich structures (yellow and green lines) exhibiting a heightened diffusivity compared to less proton rich structures (blue and purple lines).

Arrhenius trends for the superionic phase of each material, plotted as natural log of H ion diffusivity vs. 1/T, are given as the right panels of Fig.~\ref{fig:nOfR_etc}. From the slope of these lines one can infer the activation energy required for H ions to jump between adjacent sites. We see that the yellow line, representing simulation results for the highest proton fraction structure, NH$_{24}$-$C2$, has the shallowest slope and thus lowest activation energy. However, all other structures, regardless of proton fraction, have similar slopes and thus, similar activation energies for proton diffusion. This deviates from the results of de~Villa \textit{et al.} 2023\cite{devilla2023}, whose simulations of HCNO, HCN, and HNO structures showed a strong correlation between proton fraction and activation energy. These results indicates that while proton fraction is an important control on the temperature at which a material becomes superionic, it is not the only control on diffusion rate. 
The D$_0$ in Eq.~\ref{eq:Arr} is often called the frequency factor, as it is the attempt frequency for H ion jumps to adjacent sites\cite{gao2020classical}. D$_0$ depends on both the concentration of H ions and the concentration of energetically accessible lattice sites. More proton rich structures have more H ions attempting to jump. Similarly, there are likely more energetically available sites for H ions to diffuse into due to the smaller size of neighboring H ions compared to N ions, and the increased distance to neighboring N ions. As such, more proton rich structures should have a higher frequency factor, and thus a higher diffusivity at a given temperature, despite activation energies being similar across proton fractions.

\subsection{Superionic Transport Mechanisms and Pathways}

A number of local diffusion processes may lead to hydrogen superionic diffusion\cite{gao2020classical}, including but not limited to direct jumps into interstitial sites, double occupancy of existing potential wells\cite{devilla2023}, concerted diffusion, and ``paddle-wheel'' mechanisms. While the mechanisms responsible for individual H ion hops can be difficult to discern, predominant diffusion mechanisms for a given structure and temperature may be inferred from statistical visualizations, such as the trajectory plots, and H position isosurfaces. 

In Fig.~\ref{fig:msd_etc}(f) and (h) we see the circular H ion trajectory patterns around the N ions, indicating that many H ions spend significant time bonded to rotating NH$_n$ ions. However, in these plots we also see there are interconnected pathways between N ions that H ions occasionally traverse. We also observe that for many species, upon the solid-to-superionic transition, H$_n$N$_2$ species such as H$_7$N$_2$ form, as shown in Fig.~\ref{fig:SI_methods}(f) for NH$_{10}$-$C2/m$. This indicates that there are transient H$_x$N-H-NH$_y$ species, which allow the transfer of H ions between N ions. The combination of these results implies that some fraction of H ions are being transported by a paddle-wheel mechanism. In this process, H ions bond to N ions, followed by a rotation of the NH$_n$ cluster, which aligns in such a way that an H on the original NH$_n$ cluster complexes with an adjacent N ion and hops to that cluster. This proton exchange can also occur in clusters of 3 N ions, such as in H$_{10}$N$_3$, with a concerted exchange of H ions. 

From trajectory and isosurface plots we also observe more linear, direct hopping pathways available in the NH$_{10}$-$Pnma$ structure. These pathways coincide with the original positions of the H$_3^{0.5-}$ chains present at 0~K (Fig.~\ref{fig:str-xy}(c)), roughly aligned with the c-axis, with additional potential wells becoming available between the chains upon heating. 

\begin{figure*}[hbt!]
    \centering
    \includegraphics[width=1.0\linewidth]{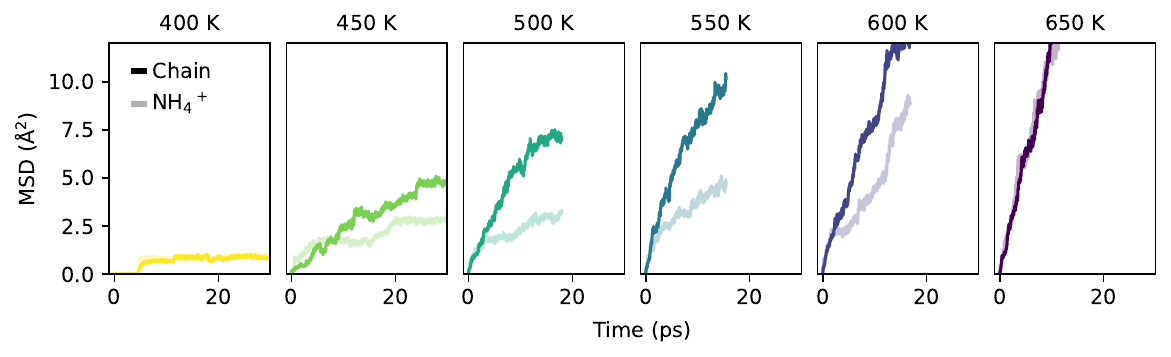}
    \caption{Comparison of the mean squared displacement of the different H ions in the NH$_{10}$-$Pnma$ structure from 400 to 650~K. The dark lines represent the behavior of for H ions that are originally located in the hydrogen chains, whereas the faint lines illustrate behavior of H ions originally in NH$_4^+$ ion. At 400~K, the material is still a solid, indicated by the flat MSD. At 450~K, the MSD increases as the material becomes superionic. From 450 to 600~K, the H ions originally on the NH$_4^+$ have lower MSDs, indicating that these H are more tightly bound and have a lower diffusion rate. (With a longer simulation time, one expects the two MSD lines to merge because exchanges between both H types occur.) At 650~K, both classifications of H ions diffuse at an equal rate indicating frequent exchanges between both reservoirs and equivalent activation energies.}
    \label{fig:chain_NH4_MSD}
\end{figure*}

Our observation of two distinct superionic diffusion mechanisms in the NH$_{10}$-$Pnma$ structure is validated by the differences observed when we sort the H ion MSDs for different initial positions. In Fig.~\ref{fig:chain_NH4_MSD}, we separately plot the MSDs for H ions originally located in the H$_3^{0.5-}$ chains, and the H ions originally in NH$_4^+$ clusters, from 400 to 650~K. (These different H ion types are shown in Fig.~\ref{fig:str-xy}(c) as yellow and white, respectively.) At 400~K, the material is still a solid, indicated by the flat MSDs. At 450~K, the MSDs increase as the material becomes superionic. From 450 to 600~K, the H ions originally on the NH$_4^+$ have lower MSDs, indicating that these H are more tightly bound,  have a lower diffusion rate, and are diffusing primarily via a different mechanism. With a longer simulation time, one expects the two MSD lines to merge because exchanges between both H types occur. At 650~K, both types of H ions diffuse at an equal rate indicating frequent exchanges between both reservoirs and equivalent activation energies. 

At higher temperatures, we observe decreasing H/N ratios (Fig.~\ref{fig:HN_ratio}, increasing lifetimes of H compared to NH$_3$ and NH$_4$ (Fig.~\ref{fig:lifetimes}, and decreasing concentrations of species with more than 1 N (Fig.~\ref{fig:SI_methods}). This suggests that as temperature increases, direct hopping mechanisms become more prevalent compared to H ion exchange through the paddle-wheel mechanism.

\subsection{Influence of Temperature, Pressure, and Proton Fraction on Species Concentrations and Lifetimes}

\begin{figure*}
    \centering
    \includegraphics[width=1\textwidth]{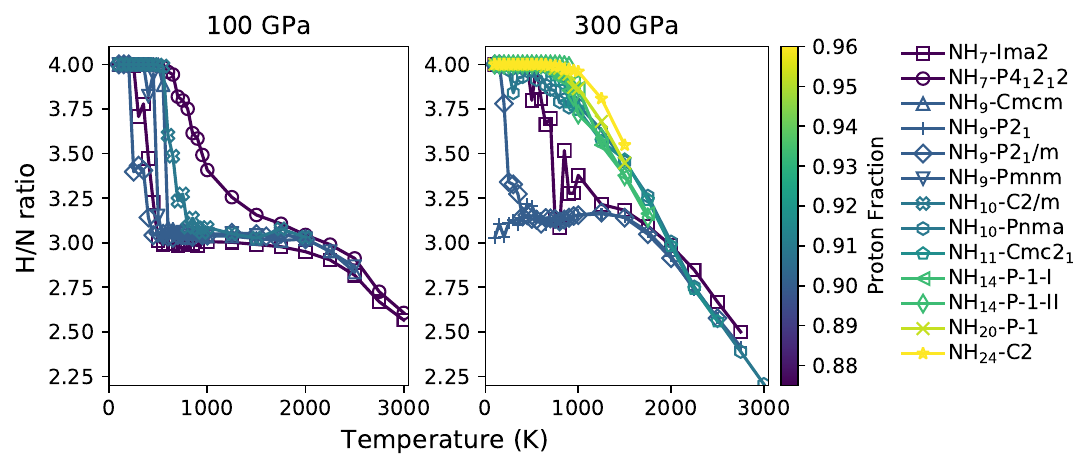}
    \caption{For all 13 compounds, H-N species have been analyzed in trajectories at different temperatures. The H/N ratio averaged over all species bearing both H and N is plotted as function of temperature for 100 and 300~GPa. The line colors reflect the proton fractions of the materials as the color bar and legend indicate. At 100~K, all materials have an H/N ratio of 4 reflecting their ammonium like nature, with the exception of the NH$_9$-$P2_1$ structure that is composed of NH$_3$ units. By $\sim$1000~K, less hydrogen rich compounds (NH$_7$-NH$_9$) transition from a NH$_4$ to NH$_3$ dominated regime as the drop of the H/N ratio to 3 indicates. More hydrogen rich structures (NH$_{10}$-NH$_{24}$) only approach a H/N ratio of 3 by $\sim$2000~K. Heating above 2000~K results in a continued decrease in the H/N ratio among the present species.}
    \label{fig:HN_ratio}
\end{figure*}

\begin{figure*}
    \centering
    \includegraphics[width=1\textwidth]{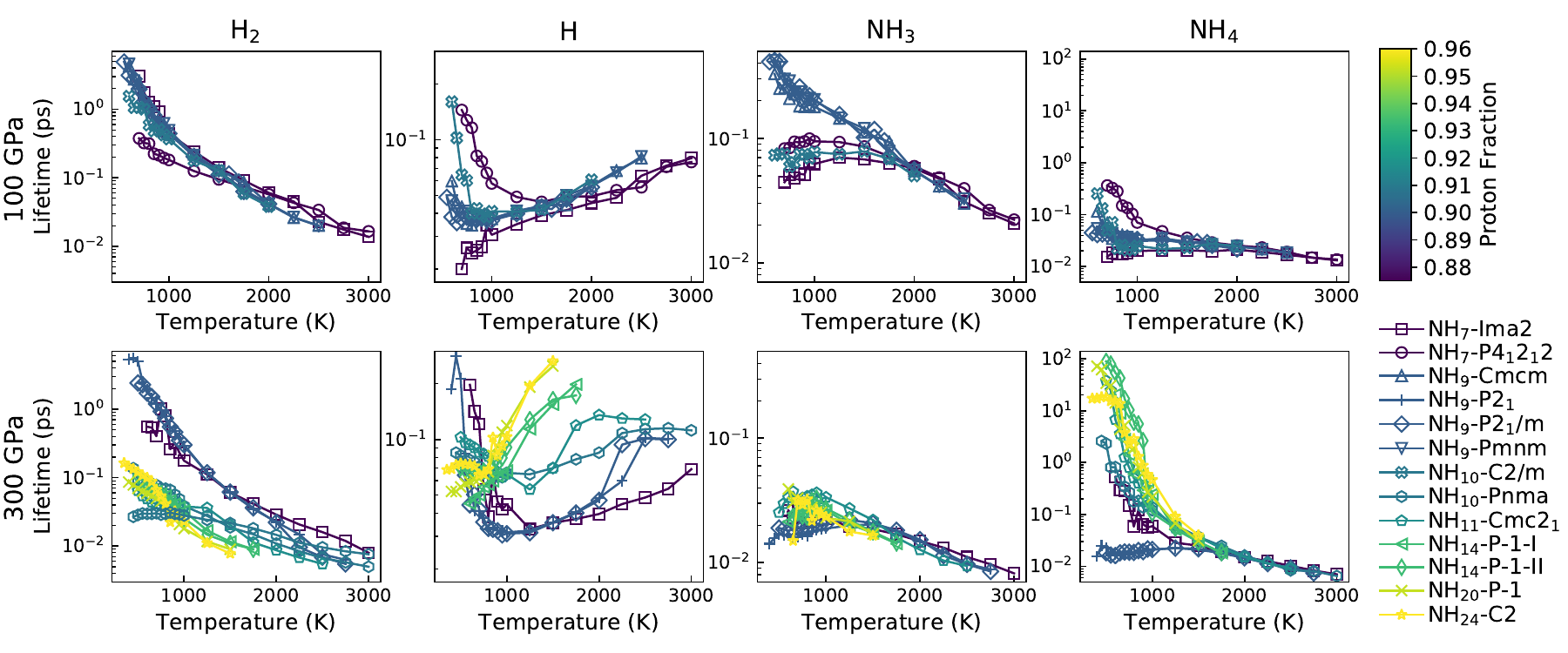}
    \caption{Average molecule lifetimes for the most prevalent species observed, H$_2$, H, NH$_3$, and NH$_4
    $, shown for 100 and 300~GPa. Temperatures shown are for the superionic and liquid phases of each compound. Several structures (NH$_{14}$-$P$-1-II, NH$_{20}$-$P$-1, and NH$_{24}$-$C2$) show no detectable NH$_3$ until 50-300~K past their solid-to-superionic transition. }
    \label{fig:lifetimes}
\end{figure*}

We analyze how chemical bonding and species lifetimes evolve with temperature by heating all structures from 100~K to their melting point and performing cluster analysis at each condition. Note that cluster analyses do not infer species charges and thus species labels in Figs.~\ref{fig:HN_ratio} and~\ref{fig:lifetimes} are merely stoichiometric indicators.

Fig.~\ref{fig:HN_ratio} shows the H/N ratios of all species which contain both H and N present at a given temperature for each initial structure, at 100 and 300~GPa. At temperatures below the solid-to-superionic transition, each structure has an H/N ratio of 4, indicative of a NH$_4$ dominated regime. (The exception is the NH$_9$-$P2_1$ structure, which as a molecular species is initially composed of NH$_3$ units, not NH$_4$.) Less proton rich structures (NH$_7$-NH$_9$) typically transition to an NH$_3$ dominated regime by 1000~K, indicated by the drop of the H/N ratio to $\sim$3. This transition is not sensitive to pressure, as structures at both 100 and 300~GPa exhibit similar H/N declines with temperature. More proton rich structures (NH$_{10}$-NH$_{24}$) only approach an H/N ratio of 3 at $\sim$2000~K, indicating that NH$_4$ is stable to higher temperatures in more proton rich structures. At temperatures beyond 2000~K, all compounds exhibit similar trends in H/N ratios, continually decreasing with temperature to values of $\sim$2-2.5 by 3000~K, corresponding to NH$_2$ and NH$_3$ dominated compositions.

We further investigate lifetimes for the most prevalent species observed throughout our simulations, H$_2$, H, NH$_3$, and NH$_4$, shown in Fig.~\ref{fig:lifetimes} for 100 and 300~GPa, with compounds colored by proton fraction as in previous figures.  Species lifetimes are shown for temperatures above the solid-to-superionic transition temperature, as at lower temperatures bonds concentrations are stable and measured lifetimes represent the duration of the simulation. 

H$_2$ is initially present in most, but not all structures studied. Regardless of pressure and proton fraction, H$_2$ lifetimes decrease with temperature, with the exception of the NH$_{10}$-$Pnma$ structure which shows a constant H$_2$ bond lifetime up to $\sim$1000~K. However, at low temperatures, NH$_{10}$-NH$_{24}$ structures exhibit H$_2$ lifetimes roughly an order of magnitude lower than NH$_7$-NH$_9$ structures. 

H lifetimes exhibit mixed trends based on pressure and proton fraction. Most compounds show a decrease in H lifetime as temperature increases to 1000-1500~K, followed by an increase in H lifetime. However, for more proton rich structures of NH$_{14}$-NH$_{24}$, H concentrations show a plateau or continuous increase with temperature. An interesting exception is that at 100~GPa, the NH$_7$-$Ima2$ structure exhibits a very low initial H lifetime at low temperatures, continually increasing, whereas at 300~GPa, this structure shows a significant decrease in H lifetime. Above $\sim$1500~K, the NH$_7$-$Ima2$ structure exhibits similar bond lifetimes for H at both pressures. At 300~GPa for temperatures above 1000~K, there is significant variation in H lifetime based on proton fraction, with more proton rich structures (NH$_{11}$-NH$_{24}$) exhibiting H lifetimes up to an order of magnitude higher than less proton structures. While a number of structures at 0~K contain hydridic H, these results show that upon heating these H ions become unstable, likely bonding into H$_2$ with H ions originally in NH$_4$ clusters, but with further heating, H ions becomes a preferred state over H$_2$. 

The comparative lifetimes of NH$_3$ and NH$_4$ diverge immensely with pressure. At 100~GPa, NH$_3$ and NH$_4$ exhibit similar bond lifetimes, ranging from 1-50~fs. NH$_9$ structures show a particularly high NH$_3$ lifetime at low temperatures. At 300~GPa, however, NH$_4$ is significantly more stable than NH$_3$ at low temperatures. NH$_3$ is not even an observable species in the NH$_{14}$-$P$-1-II, NH$_{20}$-$P$-1, and NH$_{24}$-$C2$ structures until 50-300~K past their solid-to-superionic transition, whereas NH$_4$ exhibits bond lifetimes up to 100~ps (truncated only by the end of the simulation) in high proton fraction structures at low temperatures. At high temperatures, NH$_3$ and NH$_4$ lifetimes are similar at 100 and 300~GPa. 

Our minimum bond lifetime heuristic, which is formed around the assumption that such minimum lifetimes would allow experimental observation, assumes that N-H and H-H bonds are only observable via Raman spectroscopy if they persist for 20~fs=2$\times$10$^{-2}$~ps or more. Based on this metric, at 100~GPa, H$_2$, NH$_3$, and NH$_4$ should be detectable up to $\sim$2000~K for most initial structures. For 300~GPa, temperature cutoffs for Raman detectability will be slightly lower, especially for H$_2$ in more proton rich structures (NH$_{10}$-NH$_{24}$).

\section{Conclusion}
We performed DFT-MD simulations of ammonium polyhydrides with stoichiometries ranging from NH$_7$ to NH$_{24}$ at pressures of 100-300~GPa and temperatures of 100-3000~K. By gradually heating these materials, we determined that all materials undergo a solid-to-superionic transition between 350 and 700~K. Solid-to-superionic transition temperatures depended weakly on their proton fraction and pressure. In contrast,  melting temperatures varied between 700 and 3000~K and showed a strong dependence on proton fraction. From this we conclude that NH$_n$ materials with a proton fraction greater than $\sim$0.97 are unlikely to exhibit a superionic phase, as the excess of superionic hydrogen ions would weaken the forces between N ions to the extent that no stable nitrogen sublattice could be maintained. It is remarkable that a sublattice can be stabilized for such hydrogen rich compounds as NH$_{24}$. This stabilization must be facilitated by the formation of much larger NH$_4$ ions, which mutually reinforce each other over a comparatively narrow temperature interval before the entire structure melts.

All melting temperatures are below the revised pressure-temperature profiles recently predicted for the interiors of Uranus and Neptune~\cite{militzer2025ab}, implying that ammonium polyhydrides with high hydrogen concentrations such as NH$_7$ to NH$_{24}$ likely occur in liquid form in the interiors of these planets. Conversely, the melting temperatures of water and ammonia are higher and they may occur in superionic form deep inside these planets.

By combining trajectory visualization and cluster analysis methods, we observe that H ions diffuse via multiple mechanisms simultaneously. Except for the most proton rich structure, NH$_{24}$-$C2$, we observe that all materials exhibit similar activation energies for protons to hop. This indicates that the heightened superionic diffusivity of more proton rich structures arises from additional pathways or greater frequency of H ion hopping attempts. 

Both proton fraction and pressure have a significant influence on the lifetimes of H, H$_2$, NH$_3$, and NH$_4$ clusters, particularly at temperatures below 2000~K. More proton rich structures contain clusters with much higher H/N ratios from 100-2000~K, correlating to an energetic preference for NH$_4$. By 2000~K, the clusters of all structures show a continuously decreasing H/N ratio, which approaches 2.25-2.5 by 3000~K. 

Together, these results highlight the critical role of proton fraction in governing superionic phases and provide new insights into the behavior of hydrogen rich materials under conditions relevant to ice giant interiors.


\section{Data Availability Statement}
The data that support the findings of this study are available from the corresponding author upon reasonable request.

\section{Supplementary Information}
The Supplementary Information includes visualizations of all studied structures not shown in Fig.~\ref{fig:str-xy}, as well as crystal structures in POSCAR format for all structures not previously published in Refs.\cite{wang2024superconductivity,song2019exotic}.

\begin{acknowledgments}
This work was supported by the Center for Matter at Atomic Pressures which is funded by the U.S. National Science Foundation (PHY-2020249). Computational resources at Livermore Computing and the National Energy Research Scientific Computing Center (NERSC) were used.
\end{acknowledgments}

\clearpage

\bibliography{main}
\end{document}